\documentclass[
showpacs,
amsmath,amssymb,
aps,
twocolumn,
prb,
floatfix,
nofootinbib
]{revtex4-2}
\usepackage{xcolor}
\usepackage{lipsum, babel}
\usepackage{graphicx}
\usepackage{amssymb}
\usepackage{dcolumn}
\usepackage{tabularray}
\UseTblrLibrary{booktabs}
 \allowdisplaybreaks
\begin{document}

\title{Originality of resonance and locking phenomena in SFS $\varphi_0$ Josephson junction}
\date{\today}

\author{M. Nashaat$^{1,2,*}$, and Yu. M. Shukrinov$^{1,3,\dagger}$}

\affiliation{$^1$\mbox{BLTP, JINR, Dubna, Moscow region, 141980, Russia}\\
	$^2$ \mbox{Department of Physics, Faculty of Science, Cairo University, 12613, Giza, Egypt}\\
	$^3$ \mbox{Dubna State University, Dubna, 141980, Russia}\\	
	$^{*}$ majed@sci.cu.edu.eg; $^{\dagger}$ shukrinv@theor.jinr.ru
}

\begin{abstract}
We demonstrate the realization and interplay of two ferromagnetic resonances in one SFS $\varphi_0$ Josephson junction. First resonance that is realized under microwave radiation is the famous Kittel resonance. The other is Buzdin one appearing as a result of interaction of superconducting current and ferromagnetic interlayer magnetization. Transformations of one type of resonance to another under variation of external electromagnetic radiation and the $\varphi_0$ junction parameters open an interesting way to manipulation both of them. The combined ferromagnetic resonance that exhibits the features of both resonances is demonstrated too. The coupling of the Josephson phase with the magnetization of the ferromagnetic layer, caused by the spin-orbit interaction, leads to
double synchronization, namely, synchronization of both magnetic
precession and Josephson oscillations by external radiation. The obtained results demonstrate reach physics and unique opportunities for various applications.
\end{abstract}

\maketitle

\section{Introduction}

Superconductivity and magnetism are two antagonistic phenomena but they can be united in the superconductor-ferromagnet  Josephson structures. An example of this is the $\varphi_0$ junction in which unification is based on the phase difference shift in the  current-phase relation, which depends on the values of the spin-orbit coupling  and magnetization in the ferromagnetic interlayer  \cite{buzdin08,Yokoyama2014,Bergeret2015,Konschelle2015,shukrinov-ufn22}. The phase shift leads to the ferromagnetic resonance (FMR) predicted by A. Buzdin \cite{buzdin08,konschelle09}, which is related to the excitation of the magnetization precession by the superconducting current. We call it the Buzdin resonance (BR) to distinguish from the Kittel one, which appears under external electromagnetic radiation. As we show below, the uniqueness of superconductivity and magnetism coupling, realized in the  $\varphi_0$ junction, allows one to manifest both resonances in one system. 

Additionally, two types of synchronization phenomena appears in this system: the electric component of radiation leads to the Shapiro step (SS) in the IV-characteristics and its manifestation in the magnetization dependence on the bias current \cite{sara22}; the magnetic component of radiation leads to the Buzdin step (BS) in the IV-curve and a bubble structure in magnetization reflecting the synchronization of Josephson oscillations \cite{shukrinov-prb24}. The combination of electric and magnetic fields demonstrates a chimera step in the IV-curve that shows both features of the Shapiro and Buzdin steps \cite{shukrinov-prb24, Nashaat2024}. It was stressed that the width of the chimera step is not a trivial sum of the SS and BS.   

The coupling between the superconducting phase and magnetization in Josephson junctions is intensively discussed in the materials with spin-orbit coupling \cite{Reynoso2008,Brunetti2013,Nesterov2016,Kuzmanovski2016,Assouline2019,Alidoust2021,Alidoust2021,Hasan2022,GAbobkov2024_1} and in topological insulator-based JJs \cite{Linder2010,LU2015,Schrade2015,Zyuzin2016,Schrade2017,Zhang2022}. The anomalous phase shift $\varphi_{0}$ in the current phase relation emerges from the simultaneous breaking of inversion and time-reversal symmetries \cite{shukrinov-ufn22}. As a result, a spontaneous supercurrent at zero phase difference appears, and several phenomena were predicted, e.g., magnetization reversal \cite{Shukrinov2017,Guarcello2020,Bobkova2020}, manipulating magnetization dynamics \cite{Nashaat2019,GAbobkov2024_2,GAbobkov2024_3}, Duffing effect \cite{Shukrinov2021_d}. Under ac magnetic field the IV-curve demonstrates magnetic resonance side-bands  \cite{Maekawa_SUST2011,Efetov_PRB2011} and a step structure like Devil's staircase \cite{Hikino2011,Nashaat2018}.

Experiments on FMR in superconductor - ferromagnetic (SF) heterostructures demonstrate a giant downward shift of the resonance field, which was explained by the spin-transfer torque due to spin-triplet supercurrents \cite{L_Li}. In other experiments described in Ref.\cite{Tao_Yu_2025,Jeon_2019, Golovchanskiy2020,Golovchanskiy2023}, the shift was explained in terms of the demagnetization effect \cite{Mironov_2021}. The magnetoelectric effect through Anderson-Higgs mechanism of magnon mass generation \cite{Silaev_2022} and the study of the magnetization dynamics involving the Meissner supercurrent \cite{Zhou_2023} give as well good agreement with experimental results \cite{Zhang_2025}. Recently, an alternative mechanism for the FMR shift was demonstrated experimentally and theoretically in Ref.\cite{Popadiuk2026} where different structures are explored using nonmagnetic metallic and insulating spacers. It was suggested that the magnetization precession in the F-layer modulates the magnetic flux in such structures, thereby inducing an alternating supercurrent in the S-layer, which in turn produces a dynamic back-action magnetic field on the F-layer and shifts its resonance frequency. We stress that in the mentioned papers, the FMR was considered in the absence of the Josephson effect. Below we consider the FMR in the short SFS $\varphi_{0}$ Josephson junctions taking into account magnetic component of external electromagnetic radiation which was not investigated in detail yet.

In this paper, we present the results which show the originality of the resonance and locking phenomena in the SFS $\varphi_0$ junction. A unique realization and interplay of the Kittel and Buzdin ferromagnetic resonances in one SFS-system is demonstrated. We concentrated on the influence of the magnetic component of external electromagnetic radiation, which interacts with the magnetization of the ferromagnetic layer and leads to the emergence of the Kittel, Buzdin and combined resonances. We show also the transformations of one type of resonance to another one and, additionally, the transformations of resonance to the locking and vice versa. The results are observed in two different geometries: when the magnetic component of external electromagnetic radiation is in-plane or out-of-plane to the surface of junction. An interesting situation is considered when the combined resonances are manifested in the SFS $\varphi_0$ junction.

The paper is organized as follows. The model is introduced in Section II. In section III we demonstrate the results of investigating the resonance and locking in two different geometries. Section III(A) is devoted to the manifestation of the Buzdin and Kittel resonances, while the transformation between resonance and locking is shown in III(B). In section III(C) we show a clear manifestation of combined resonances. Finally, we come to the conclusions in section IV.
	
 \section{Model}
As we have mentioned above, the  SFS $\varphi_0$ junction is characterized by the direct coupling between the superconducting phase difference across the junction $\varphi$ and the magnetization $\mathbf{M}$ of the ferromagnetic layer \cite{buzdin08}. Its dynamics can be described by the system of equations including the Landau-Lifshitz-Gilbert (LLG) equation, the equation for the biased current of the resistively and capacitively shunted junction (RCSJ) model and the Josephson relation for the phase difference and voltage \cite{konschelle09,shukrinov-prb19}.

The Landau-Lifshitz-Gilbert (LLG) equation has a form:
\begin{eqnarray}
\frac{d\mathbf{M}}{dt}=&-\gamma \mathbf{M}\times \mathbf{H}_{eff}+\frac{\alpha }{M_s}\bigg( \mathbf{M}\times \frac{d\mathbf{M}}{dt}\bigg),
		\label{LLG}
\end{eqnarray}
where $\gamma$ is the gyromagnetic ratio, $\alpha$ is the Gilbert damping constant and $M_s = ||\textbf{M}||$.
The effective magnetic field acting on the magnetization $\mathbf{M}$ is given by
\begin{eqnarray}
\textbf{H}_{eff}=-\frac{1}{\mathcal{V}}\frac{\delta F}{\delta \textbf{M}}
\label{HeffF}
\end{eqnarray}
where $F$ is the free energy of the system, and $\mathcal{V}$ is the volume of the ferromagnet. In this work, the free energy consists of three parts
\begin{equation}
F=E_{S}+E_{M}+E_R,
\label{F}
\end{equation}
where $E_S$ is the superconducting energy given as \cite{konschelle09}:
\begin{equation}
E_S = E_J \left[ 1 - \cos\left(\varphi-r\frac{M_y}{M_s}\right)\right],
\label{E_S}
\end{equation}
where $E_J = \Phi_{0}I_c/2\pi$ is the Josephson energy, and $I_{c}$ is the critical current of the junctions, and $\Phi_{0}=h/2e$ is the flux quantum. The relative strength of spin-orbit coupling is characterized by the Rashba type parameter
$r$.

In the proposed junction, we assume the ferromagnet has an easy axis along z-direction, hence the magnetic anisotropy energy $E_{M}$ reads as 
\begin{equation}
E_M = - \frac{K_{an}\mathcal{V}}{2} \left(\frac{M_z}{M_s}\right)^2,
\label{E_M}
\end{equation}
where $K_{an}$ is an anisotropy constant. In  permalloy, doped with $Pt$~\cite{hrab16}, in the ferromagnets $MnSi$ and $FeGe$, the value of $r$ can vary in the range $0.1-1$~\cite{greening2020} with weak magnetic anisotropy $K_{an} \sim 500 \ J/m^{3}$~\cite{rus04}, and $M_{s} =860 \ A/m$. The ferromagnetic resonance frequency $\Omega_{F}=\gamma K_{an}/M_{s}=16.3$ GHz. Finally, $E_R$ represents the energy related to the magnetic component of the external radiation with amplitude $H_{R}$ given as
\begin{equation}
E_R = - (\mathbf{M},\mathbf{H}_R),
\label{E_R}
\end{equation}
where
\begin{equation}
\textbf{H}_{R} =  H_{R}\sin(\Omega_{R}t)\hat{\textbf{e}}_{y(z)},
\end{equation}
where $\Omega_{R}$ is the radiation frequency, $\hat{\textbf{e}}_{y(z)}$ indicates two considered geometries (see below).

\begin{figure}[h!]
 	\centering
 	\includegraphics[width=0.6\linewidth]{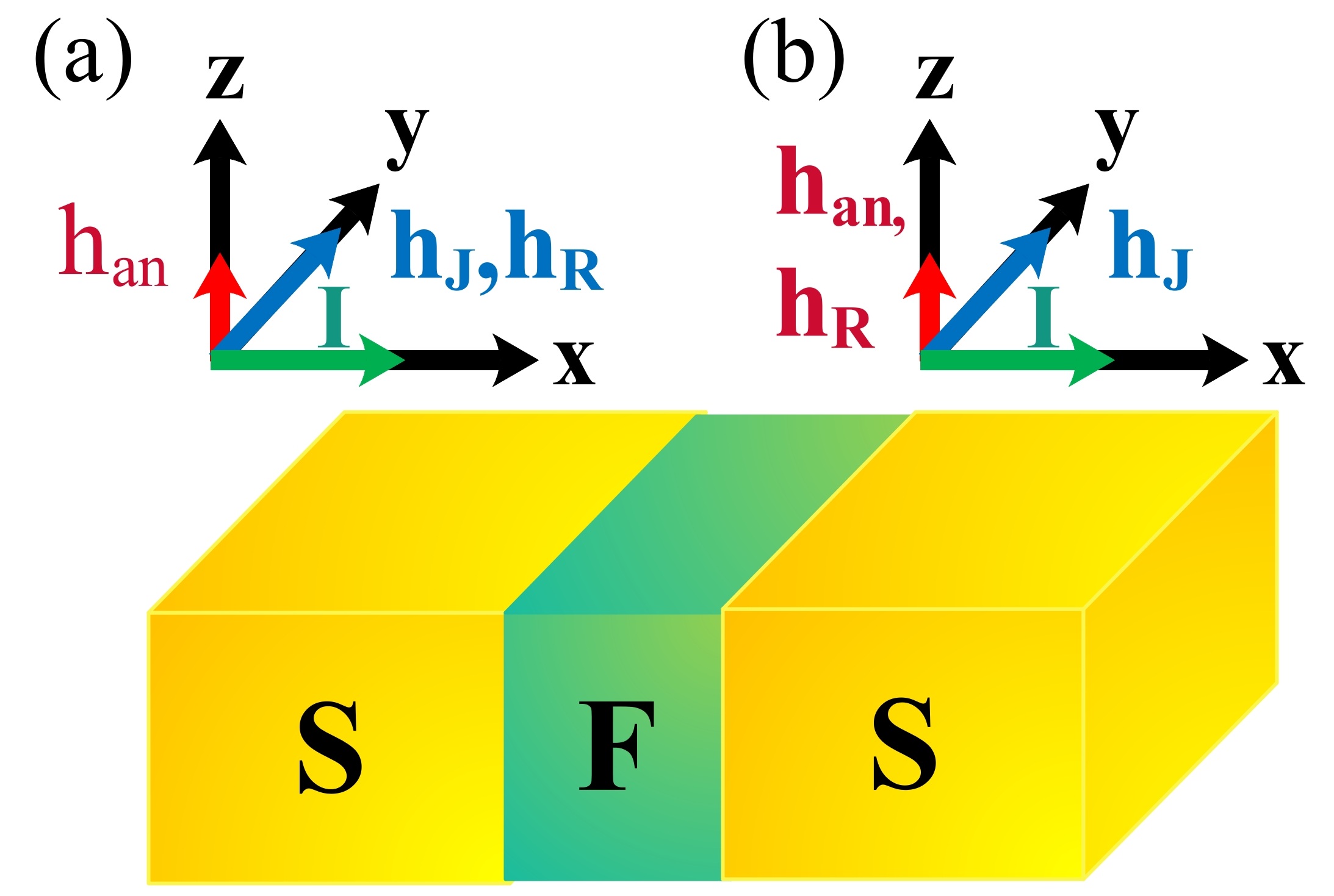}
 	\caption{(a) In-plane geometry: effective  field due to the magnetic component of radiation $h_R$ is parallel to the Josephson effective field $h_J$ and perpendicular to the anisotropy field. (b) Off-plane geometry: $h_R$ is perpendicular to $h_J$.}
 	\label{1}
 \end{figure}

We investigate the originality of resonance and locking phenomena in $\varphi_0$ junction in two distinct geometries with the ferromagnet easy-axis along the z-axis: when the radiation effective  field is aligned parallel to the Josephson effective field and directed along the y-axis (G1-geometry, see Fig.\ref{1}(a)), and when it is oriented perpendicularly to the Josephson effective field (G2-geometry, see Fig.\ref{1}(b)). 
 In both geometries the effective field component along x-axis is zero. The complete system which includes the LLG and RCSJ equations in the dimensionless form for the dynamics of the $\varphi_0$ junction is given by \cite{Guarcello2020,shukrinov-prb24}: 
\begin{eqnarray}
	\frac{d\mathbf{m}}{dt}&=&- \frac{\omega_{F}}{1+\alpha^{2}}\bigg(\mathbf{m}\times \mathbf{h}_{eff}+\alpha [\mathbf{m}\times\left( \mathbf{m}\times  \mathbf{h}_{eff}\right)]\bigg), \nonumber \\
	\dot{V}&=& \left[ I-V+r\dot{m}_{y}-\sin (\varphi -\varphi_{0})\right]/\beta_c, 
	\label{LLGjj}
\end{eqnarray}
where the voltage is determined by the Josephson relation $\dot{\varphi}=V$, $\beta_c$ is the McCumber parameter, $m_{y}$ is the magnetization component along y-direction normalized to the saturation magnetization $M_{s}$. The effective field $h_{eff}$ is scaled to the magnetic anistropy energy. The bias current $I$ is normalized to junction critical current $I_{c}$. The sine function has the additional phase shift $\varphi _0=rm_y$ \cite{buzdin08,shukrinov-prb24}. Time is normalized to $\omega _c^{-1}$, where $\omega _c=2eI_cR_{N}/\hbar $ is the characteristic frequency of the junction, and $R_{N}$ is the junction resistance in the normal state. The ferromagnetic resonance frequency $\Omega _F$ is scaled to $\omega _c$, such that $ \omega_{F}=\Omega_{F}/\omega_{c}$. The average voltage $V$ across the junction is normalized to $v_{c}=I_{c}R_{N}$. According to the given normalization, we have $\omega_{J}=V$. The effective field in G1-geometry is given by
\begin{eqnarray}
 h_{y} =G r \sin(\varphi - r m_{y})+  h_{R} \sin(\omega_R t),  \ \ \  h_{z} = m_{z}.
	\label{G1-effective}
\end{eqnarray}
While in G2-geometry it reads as
\begin{eqnarray}
	h_{y} =G r \sin(\varphi - r m_{y}), \ \ \ h_{z} = m_{z} +h_{R} \sin(\omega_R t). 
	\label{G2-effective}
\end{eqnarray}

where the frequency of external radiation $\Omega _R$ is  scaled to $\omega _c$, such that $\omega_{R}=\Omega_R/\omega_c$. The normalized amplitude of the magnetic component is given by $h_R=\frac{\gamma }{\omega _c}H_R$. The magnetization components are scaled to the saturation value, so $m_{i}=\frac{M_{i}}{M_s}$. The ratio of the Josephson and magnetic anisotropy energies is represented by ${G=E_J/(K_{an}\mathcal{V})}$. The value of the product $Gr$ can be in the range of $0.01-100$~\cite{konschelle09, shukrinov-prb24}. Depending of the critical current and junction resistance in normal state,  the characteristic frequency $\omega_{c}$ can be $30$ GHz for junction with critical current density $J_{c}$ = $10^{5}$ A/cm$^{2}$ and area (0.1 × 0.1) $\mu m^{2}$. For  $H_{R}$ range from $10 \mu$T to 53.5 mT, the value of $h_{R}$ takes values between $10^{-5}$ to 0.05.  We stress that there is an upper limit of the Josephson to magnetic energy ratio (G value). In this limit a complete reorientation of the easy-axis might occur depending on the Josephson frequency, Gilbert damping and spin-orbit coupling \cite {Shukrinovepl2018}. In the current work we consider small value of $G$ such that the resonance features are clearly manifested in the IV-curve and magnetization of the F-layer.

\begin{figure*}[t!]
	\center{
		\includegraphics[width=0.48\linewidth]{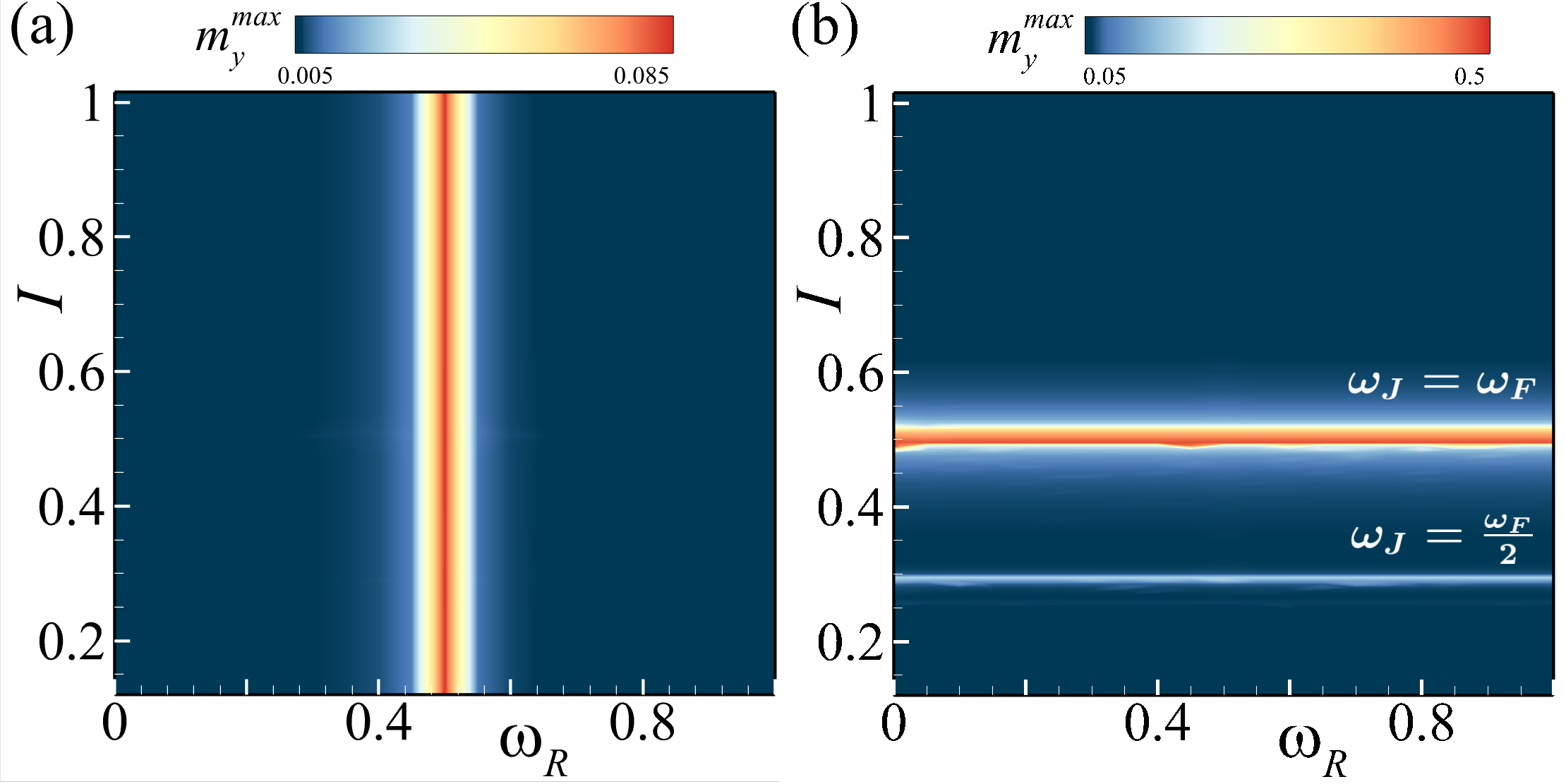}
		\includegraphics[width=0.45\linewidth]{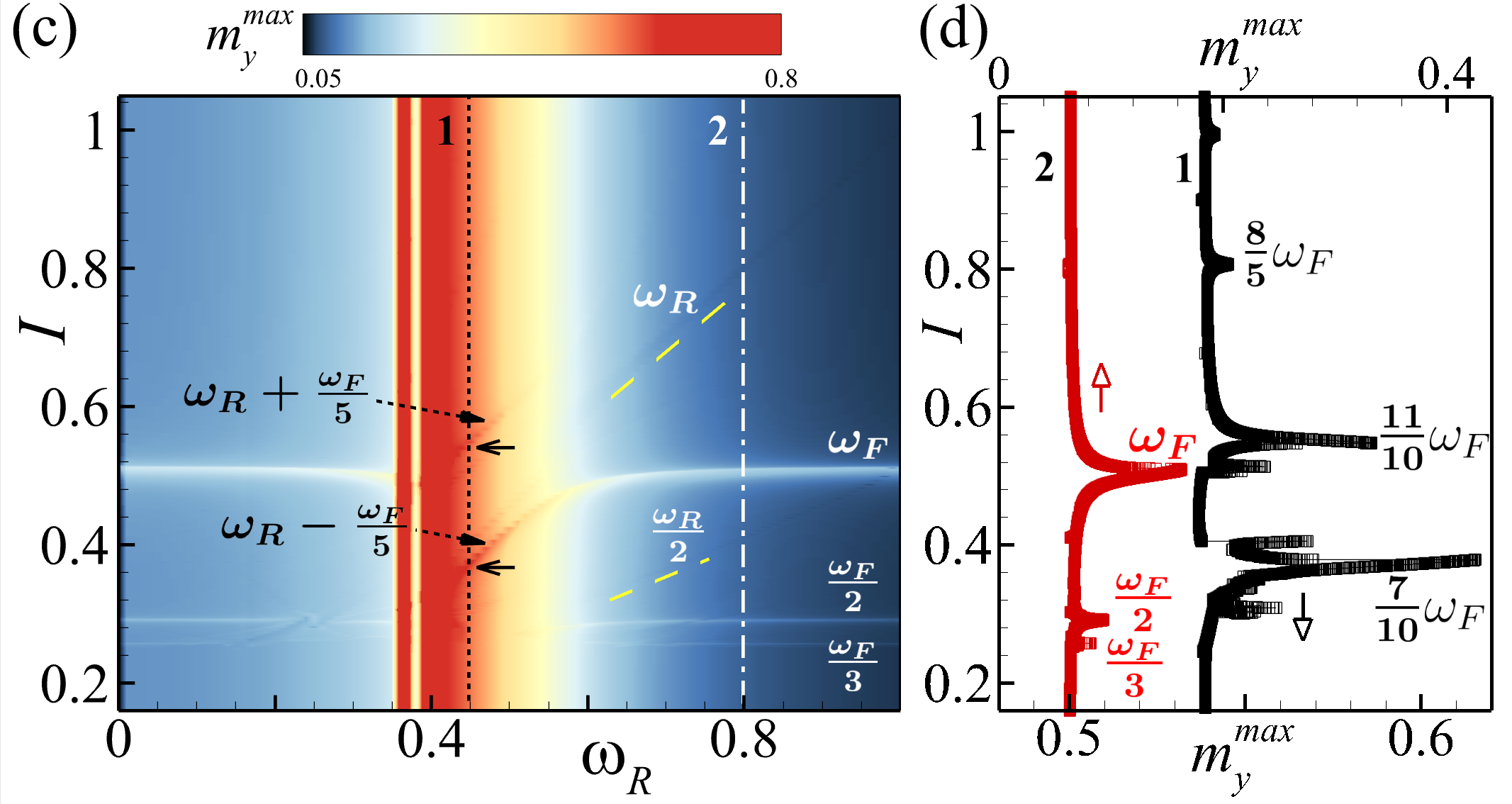}}
	\caption{Manifestation of different resonances on $I-\omega_R$ diagram. (a) Kittel resonance at $G=10^{-4}$, $r=0.1$ and $h_{R}=0.001$. (b) Buzdin resonance at $G=0.05$, $r=0.2$ and $h_{R}=10^{-5}$. The combined resonance shown in (c) at  $G=0.01$, $r=0.2$ and $h_{R}=0.05$. (d) The $I$-dependence of $m_{y}^{max}(I)$ near Buzdin - Kittel resonance region (line 1), and demonstrating the combined resonance peaks (see plain arrows) and far from the KR-region (line 2), demonstrating Buzdin resonances and its subharmonic peaks. }	
	\label{fig2}
\end{figure*}

The system of equations (\ref{LLGjj}) and (\ref{G1-effective}) for G1 geometry, and (\ref{LLGjj}) and (\ref{G2-effective}) for G2 geometry is solved numerically using the fourth-order Runge-Kutta method. It yields $m_i(t)$, $V(t)$, and $\varphi(t)$ as functions of the external bias current $I$. After using the averaging procedure \cite{shukrinov2007,buckel2008}, we can find the IV-characteristic at the fixed system parameters. All presented results are obtained at $\omega_{F}=0.5$, $\alpha=0.01$, $r=0.2$ and $G=0.01$, if it does not mentioned in the text. Dynamics of the system is analyzed in cases when only the magnetic component of radiation affect the system. 

 \section{Results}
 \subsection{Interplay of the Kittel and Buzdin Resonances}
In the G1-geometry, the magnetic moment experiences a torque from the Josephson effective field $G r \sin(\varphi - r m_{y})$ and the Buzdin resonance occurs when the Josephson frequency matches the resonance frequency $\omega_{F}$. On the other hand, the radiation effective field $h_{R} \sin(\omega_R t)$ induces the Kittel resonance when the frequency of the ac magnetic field is approaching $\omega_{F}$. At the same time, a locking might be possible due to the indirect acting of radiation magnetic component on the Josephson oscillations coupled with magnetization through spin-orbit coupling. This locking leads to the Buzdin steps in the IV-curve \cite{shukrinov-prb24, Nashaat2024, Kulikov2025}.  

\begin{figure}[h!]
	\centering
	\includegraphics[width=0.76\linewidth]{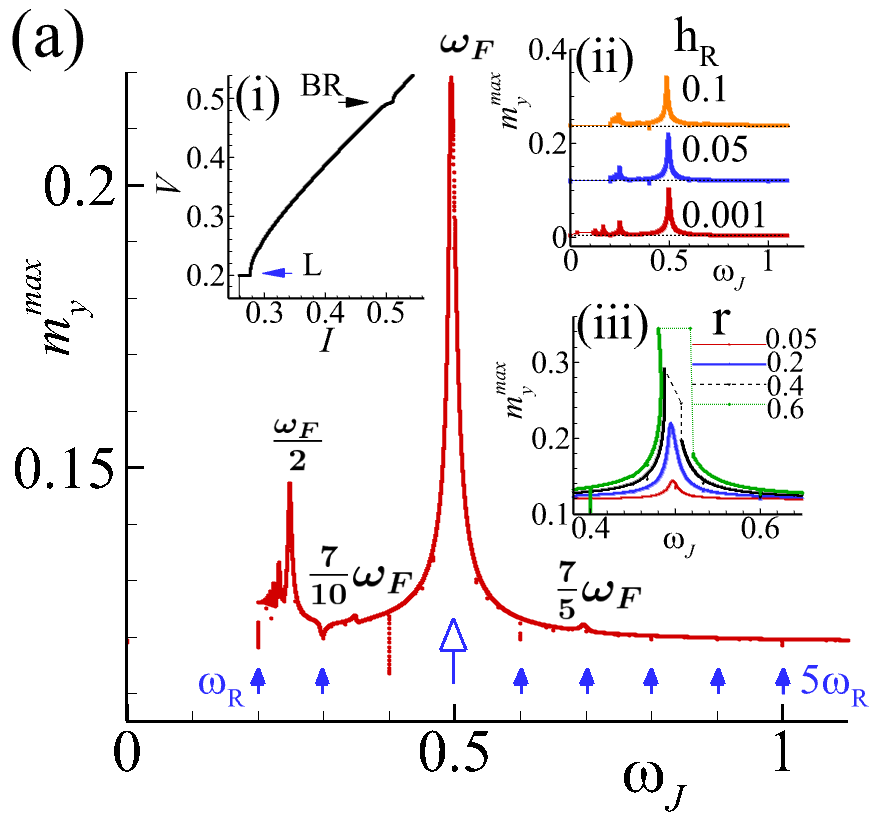}
	\includegraphics[width=0.78\linewidth]{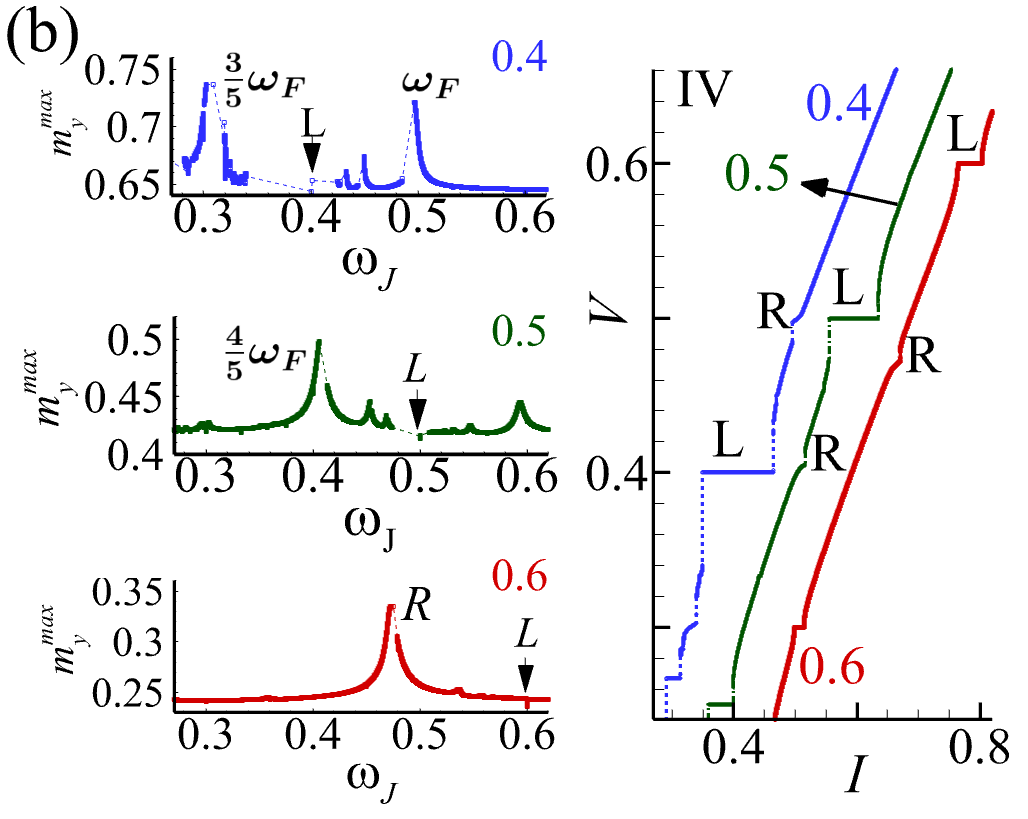}
	\caption{(a) $m_y^{max}(\omega_{J})$ in G1 at $\omega_{R}$=0.2 and $r=0.2$. Hollow arrow shows the main BR, blue filled arrows   demonstrate the manifestation of locking. Inset (i) shows the IV-curve, and insets (ii) and (iii) shows the effect of $h_{R}$ and $r$, respectively.  (b) Left side shows $m_y^{max}(\omega_{J})$ at $\omega_{R}=0.4, 0.5$ and $0.6$ with $h_{R}=0.05$. Right side shows the corresponding IV-curves. Letter "L" indicates locking, while letter "R" indicates the resonance. The IV-curves for $\omega_{R}=0.5$ and $0.6$ are shifted by $\Delta I=0.09$ to the right relatively to $0.4$ for clarity.}
	\label{fig3}
\end{figure}

Manifestation of the Kittel and Buzdin resonances in the $\varphi_{0}$ JJ by variation of model parameters is illustrated by Fig.\ref{fig2}. A sharp symmetric line emerging vertically at $\omega_{R}=\omega_{F}=0.5$ in Fig.\ref{fig2}(a) manifests the Kittel resonance (KR). By changing the ratio of Josephson to magnetic energy \textit{G}, the spin-orbit coupling \textit{r} and the amplitude of the external radiation \textit{$h_{R}$}, we observe two horizontal resonance lines $\omega_{J}=\omega_{F}$ and $\omega_{J}=\omega_{F}/2$ in Fig.\ref{fig2}(b) correspond to harmonic and subharmonic of Buzdin resonance. These two resonances and their combination is demonstrated in Fig.\ref{fig2}(c). For the simulated parameters, the KR occurs at $\omega_{R} \approx 0.4$ due to the nonlinearity of the LLG and the magnetic anisotropy. Two inclined lines (marked by dashed arrows) emerge within the intersection region of the BR and KR, indicating combined resonances at $\omega_{R} \pm \omega_{F}/5$. Additionally, the faint lines correspond to $\omega_{J} = \omega_{R}$ and $\omega_{J} = \omega_{R}/2$ (dashed traces) are related to the locking, i.e., the manifestation of the Buzdin step and its subharmonic \cite{shukrinov-prb24}. Whether BR or KR dominates the spectrum is dictated by the relative magnitudes of $h_{R}$ and $G$ \cite{shukrinov-prb24}. 

 This interplay of resonances is demonstrated in Fig.\ref{fig2}(d), where the corresponding $m_{y}^{max}(I)$ dependence are shown. We see that the KR dominates in the first region at $\omega_{R}=0.45$ (dotted line 1), while the BR dominates  in the second region at $\omega_{R}=0.8$ (dash-dotted line 2). The $m_{y}^{max}(I)$ curve at $\omega_{R}=0.45$ intersects the two lines of combined resonances (see plain arrows in (c)), yielding two main resonance peaks in $m_{y}^{max}(I)$ dependence (see black curve, line 1) at $\omega_{J}=7/10 \ \omega_{F}$ and $\omega_{J}=11/10 \  \omega_{F}$, satisfying the conditions $\omega_{J}=\omega_{R}-\omega_{F}/5$ and  $\omega_{J}=\omega_{R}+\omega_{F}/5$, respectively. A third peak corresponded to $8/5 \ \omega_{F}$ satisfies the condition $\omega_{J}=\omega_{R}+7\omega_{F}/10$. 
 
 At $\omega_{R}=0.8$, where the BR dominates, the prominent features shown by red curve (line 2) in $m_{y}^{max}(I)$ dependence include BR peak at $\omega_{J}=\omega_{F}$ and its subharmonics at $\omega_{J}=\omega_{F}/2$ and $\omega_{J}=\omega_{F}/3$.

 The G1 geometry reveals a distinct spectrum of resonant features. The two main resonances arise from the direct driving of the magnetization precession by ac effective fields: Kittel resonance by the radiation field and Buzdin resonance by the Josephson effective field. The $\varphi_{0}$ JJ is a mixed state of superconductivity and magnetism related to the coupling of Josephson phase difference and magnetization of the F-layer. This coupling leads to the combined resonance. In this regime, the system responds not only to the fundamental frequencies, $\omega_{R}$ and $\omega_{J}$, but also to their combinations. As illustrated in Fig. \ref{fig2}, a pronounced resonance peak occurs when any of these frequency combination coincides with the natural ferromagnetic resonance frequency $\omega_F$.
 
\begin{figure*}[t!]
	\center{
		\includegraphics[width=0.3745\linewidth]{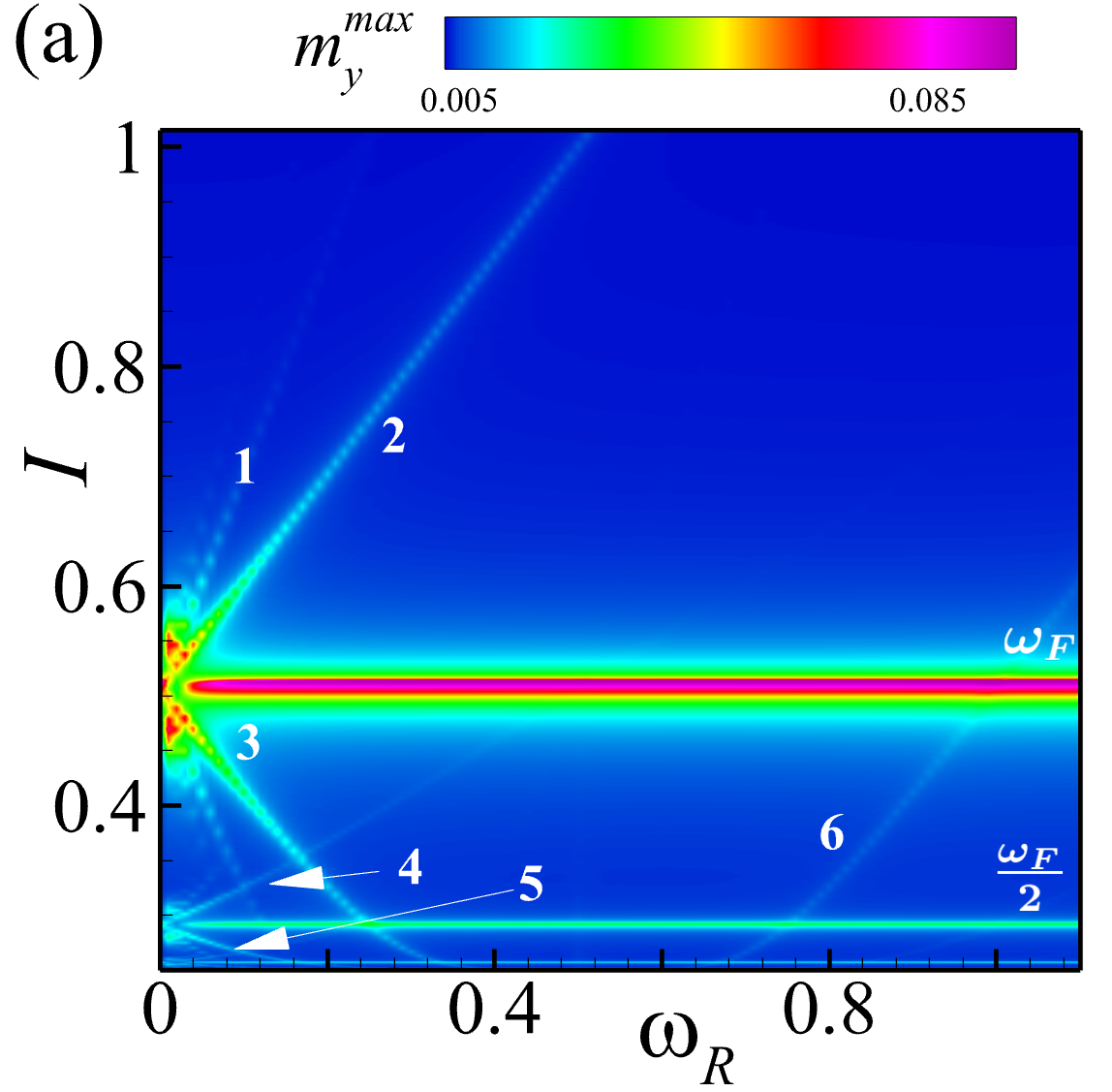}
		\includegraphics[width=0.5\linewidth]{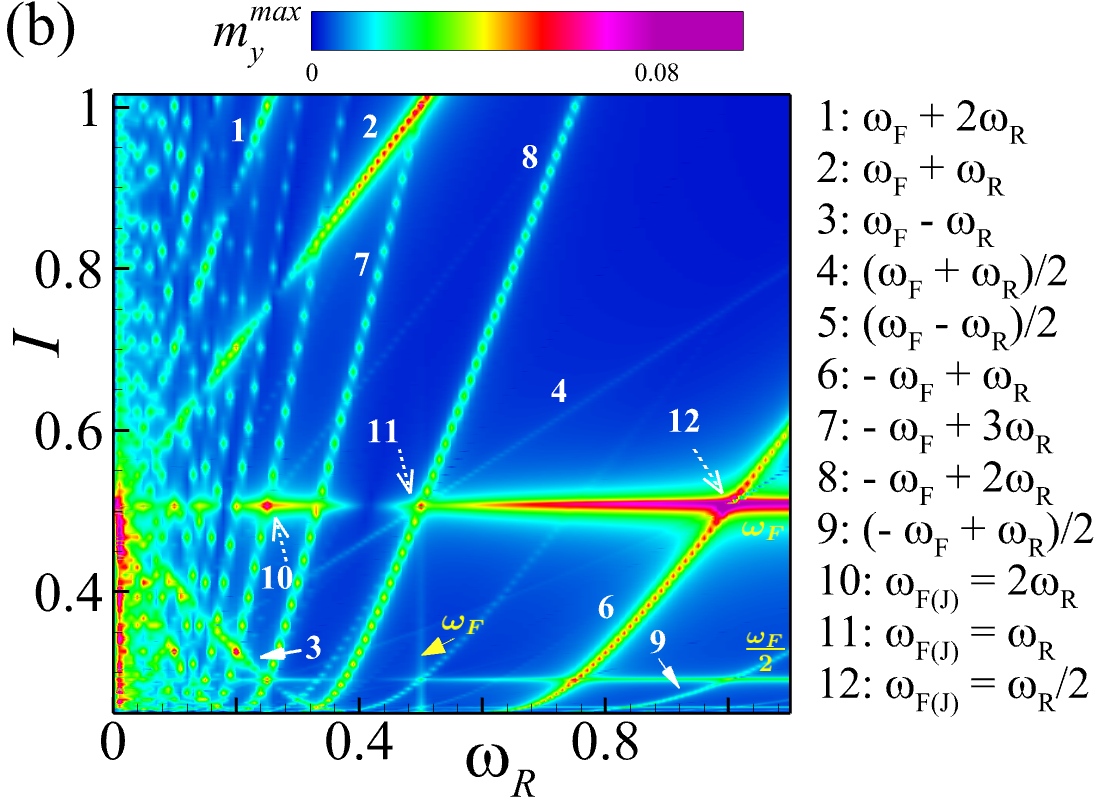}}
	\caption{(a) $I-\omega_R$ diagram of resonance maxima under radiation at $h_{R}=0.05$ in G2-geometry; (b) The same for $h_{R}=1$. Color demonstrates the value of  $m_y^{max}$ in the corresponding resonance peak. The arrows and numbers show the corresponding peak positions indicated on the right. }
	\label{fig4}
\end{figure*}

\subsection{Transformation of Resonances and Locking}

Another interesting phenomenon that can be observed in $\varphi_{0}$ JJ is the transformation of the resonance to the locking and vice versa. In Fig.\ref{fig3}(a), the $m_{y}^{max}$ as a function of $\omega_{J}$ is plotted at $\omega_{R} = 0.2$. We see a set of locking lines which appears at $\omega_{J}=n/2 \ \omega_{R}$ with integer n=2,3,... The locking at n=5 which should correspond to $\omega_{J}=2.5\ \omega_{R}$ is washed out by the Buzdin resonance peak (shown by hollow arrow). The IV-characteristic has a resonance branch, but it does not show a step at this value of $\omega_{J}$ (see inset (i)).

The locking lines $\omega_{J}=n/2 \ \omega_{R}$ indicate the indirect locking of the magnetization precession. The term  "indirect"  is used here to stress that the magnetization precession is coupled to the Josephson oscillations, and the external ac magnetic field locks the magnetization precession due to this coupling \cite{shukrinov-prb24}.

The effect of $h_{R}$ on the BR peak is demonstrated by inset (ii). It shows that its height is not affected by $h_{R}$ for the given parameters. With increasing of spin-orbit coupling, a foldover effect and nonlinear features can be manifested in the resonance peak (see inset (iii)) \cite{shukrinov-bjnano-22}. 

As we mention before the $\varphi_{0}$ junction under external radiation allows the transformation of BR to the locking of the Josephson and magnetization oscillations. In Fig.\ref{fig3}(b), we show the  $m_{y}^{max}(\omega_{J})$ dependence at three values of $\omega_{R}=0.4,0.5$ and $0.6$ around the BR condition ($\omega_{J}$ close to $\omega_{F}=0.5$). The corresponding regions in the IV-curve are demonstrated as well. The main harmonic of Buzdin step appears at $V=\omega_{R}$ (noted by letter L). Its width is larger near the resonance, and as we see, the $m_{y}^{max}$ has the largest value near $\omega_{R}=0.4$. We stress that the dependence of the Buzdin step width on the magnetic component amplitude follows a different Bessel behavior compared to the Shapiro step, and also the Buzdin step becomes more pronounced near the resonance condition \cite{Kulikov2025}. Since the IV-curve exhibits abrupt jumps near resonance and locking regions at voltages $V$ close to 0.5, 0.4 and 0.3, the corresponding resonance peaks appear truncated asymmetrically, manifested only on one side. At $\omega_{R}=0.5$ as we can see in this figure, the BR peak is washed out by locking, while it is restored at $\omega_{R}=0.6$. 

Moreover, the Kittel resonance is also manifested on the baseline value of $m_{y}^{max}$. At $\omega_{R}=0.2$, the baseline is around 0.12, which increases till 0.65 for $\omega_{R}=0.4$, then decreases till 0.25 at $\omega_{R}=0.6$. Description and analysis of the peaks demonstrated in $m_{y}^{max}(\omega_{J})$ dependence are presented in the supplemental material \cite{supplemental}.  

\subsection{Combined Resonances}
Clear manifestation of combined resonances can be obtained in G2-geometry. In this geometry the Josephson effective field is separated from the external radiation one (see Eq.\ref{G2-effective}), so we expect the appearance of Buzdin resonance. However, the magnetic component of radiation is along the easy axis now, so this affects the appearance of Kittel resonance. 

To clarify the obtained resonance features in G2-geometry, we demonstrate the $I-\omega_{R}$ diagram at two different values of $h_{R}$ in Fig.\ref{fig4}. At $h_{R}=0.05$ (see Fig.\ref{fig4}(a)) a set of horizontal lines appears at $\omega_{J}=\omega_{F}/n$ with n =1,2. Along with those lines, we observe another group of inclined lines that start to grow from the same positions at $\omega_{R}=0$. Those lines are symmetric around the starting point and satisfy the relation $(\pm m\omega_{R}+ \omega_{F})/n$, where m=1,2,3, while n=1 for $\omega_{J}=\omega_{F}$, n=2 for $\omega_{J}=\omega_{F}/2$, etc. Qualitatively, the combined resonance conditions can be found using perturbation method if one assume zero Gilbert damping and $m_{y}(t), m_{x}(t)<<1$ (see \cite{supplemental}). In this case, the zero order and $1^{st}$ order corrections for $m_{+}(t)=m_{x}(t)+im_{y}(t)$ have the following form

\begin{eqnarray}
	m^{0}_{+}(t) &=&\sum_{m,n=-\infty}^{\infty} \eta_{1}  \bigg[ \frac{e^{i ( (m+n) \omega_R + \omega_{J}) t} - 1}{   \omega_F+ n \omega_R + \omega_{J}} \nonumber \\&
-& \frac{e^{i ( (m+n) \omega_R - \omega_{J}) t} - 1}{   \omega_F + n \omega_R - \omega_{J}}\bigg]  \label{eqm0} \\
m_+^{1}(t) &=&  \sum_{m,n,P,Q} \dfrac{ \eta_{2} [e^{i ((m+n+P) \omega_R+Q\omega_{J}) t}-1] }{ \omega_F+(P+n)\omega_{R}+Q\omega_{J}}
\end{eqnarray}
	
	where $\eta_{1}=	\frac{G r}{2i} (-1)^n i^{m+n} J_m\left(\frac{\Delta \omega}{\omega_R}\right) J_n\left(\frac{\Delta \omega}{\omega_R}\right) $ and $\eta_{2}= (-1)^n i^{m+n} \Gamma J_m\left(\frac{\Delta \omega}{\omega_R}\right) J_n\left(\frac{\Delta \omega}{\omega_R}\right) $, $J_{m(n)}$  are Bessel functions of the first kind. The resonance conditions are $\omega_F = n\omega_R \pm \omega_{J}$, $\omega_{F} = P'\omega_{R} + Q\omega_{J}$, where $P'$ is integer and $Q=0,\pm1,\pm3$. By considering higher order corrections, one can generate all multiples of $\omega_{J}$, and, in general, we have $\omega_{R}=s\omega_{F}+s'\omega_{J}$, where s and s' are rational numbers.

An increase in  $h_{R}$ enhances the combined resonances. For example, at $h_{R}=1$ ( see Fig.\ref{fig4}(b)) a set of resonances related to the inclined lines appear at $\omega_J=\pm n \omega_{R} \pm \omega_F$. Another group follows the sequence $(\omega_{R}-\omega_F)/n$. When two resonance lines intersect, a new resonance appears with $n\omega_J/m$. These results are in agreement with the fact that the presence of the magnetic component and the Josephson effective fields produces a torque on the magnetization. When the resonance occurs, the torque compensates the damping term and excites the system. Finally, a vertical faint line at $\omega_{R}=\omega_F$ corresponds to the manifestation of the Kittel resonance. 
\begin{figure}[h!]
	\centering
	\includegraphics[width=0.8\linewidth]{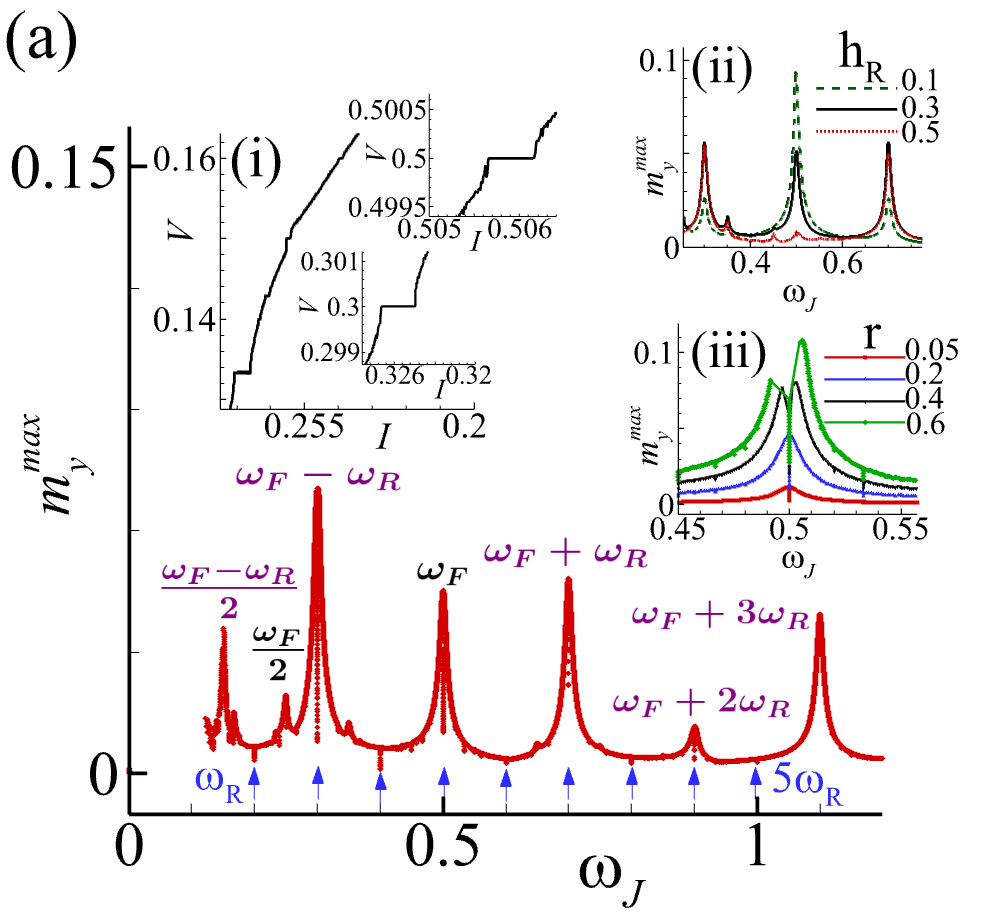}
	\includegraphics[width=0.8\linewidth]{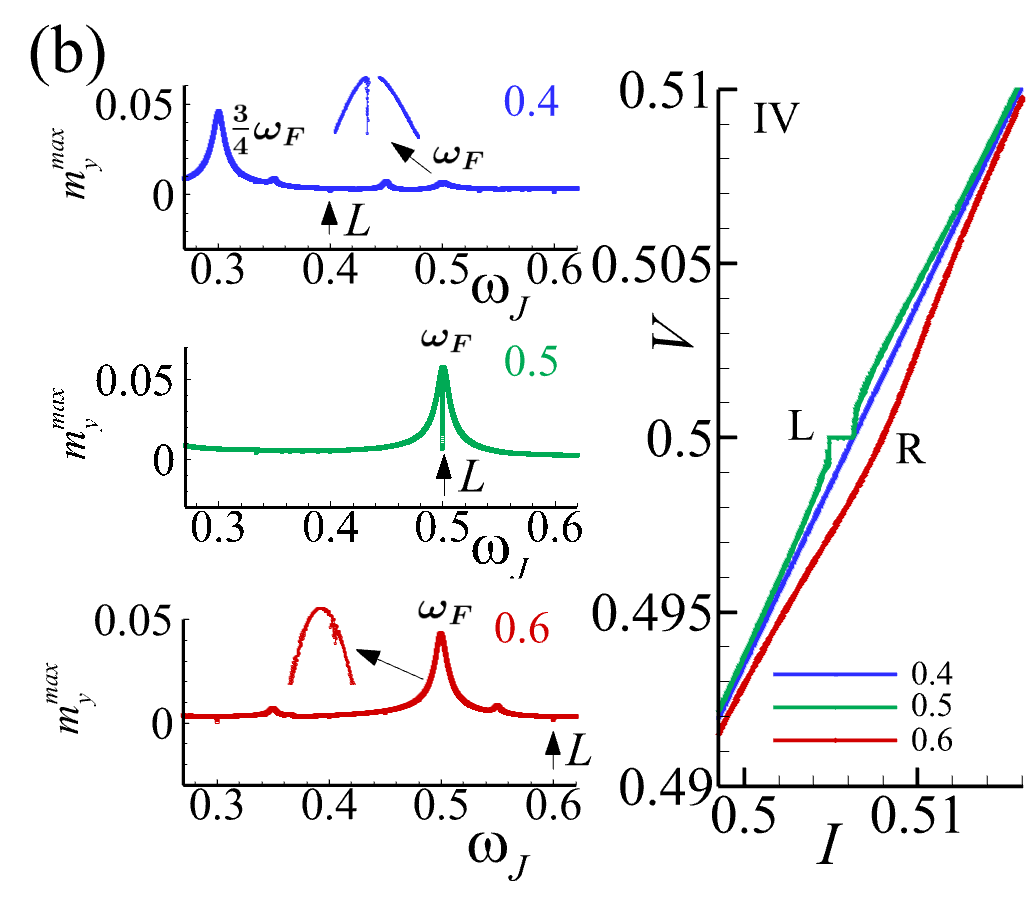}
	\caption{Interplay of resonance and locking in G2-geometry. (a) $m_y^{max}(\omega_{J})$ at $\omega_{R}$=0.2. The blue filled arrows   demonstrate the manifestation of locking. Inset (i) shows the IV-curve, and insets (ii) and (iii) shows the effect of $h_{R}$ and $r$, respectively. (b) Left side shows $m_y^{max}(\omega_{J})$ at $\omega_{R}=0.4, 0.5$ and $0.6$ with $h_{R}=1$. Right side shows the corresponding IV-curves. Letter "L" indicates locking, while letter "R" indicates the resonance.}
	\label{fig5}
\end{figure}
 A significant difference with G1 geometry is the appearance of enhanced combined resonance peaks determined by the ratio of $\omega_{F}$ and $\omega_{R}$. In addition to this, the resonance peaks occur along with the locking of Josephson and magnetization precession. In Fig.\ref{fig5}(a), we demonstrate the resonance curve at $\omega_R=0.2$. At locking conditions $\omega_{J}=n \omega_{R}/m$ shown by arrows in Fig.\ref{fig5}(a) (inset i), the Buzdin steps appear in the IV-curve. Since both anistopy and magnetic component fields are in the same direction, changing $h_{R}$ (as shown in inset (ii)) changes the BR peak at $\omega_{J}=\omega_{F}$. By increasing, $h_{R}$ the BR is getting weaker due to the strong fields along the z-direction. The torque generated by the Josephson effective field is not enough to compensate the damping part in the LLG equation. The effect of spin-orbit coupling is shown in inset (iii) and a strong splitting of the resonance peak appears due to the locking to the external radiation. We note that the foldover effect is not manifested here, as it was in G1 for the given parameters. The sidebands with $\pm n \omega_{R}\pm \omega_{F}$ appear in the $m_{y}^{max}(\omega_{J})$ dependence and a combined resonance appears together with locking. 
\begin{figure}[h!]
	\centering
	\includegraphics[width=0.6\linewidth]{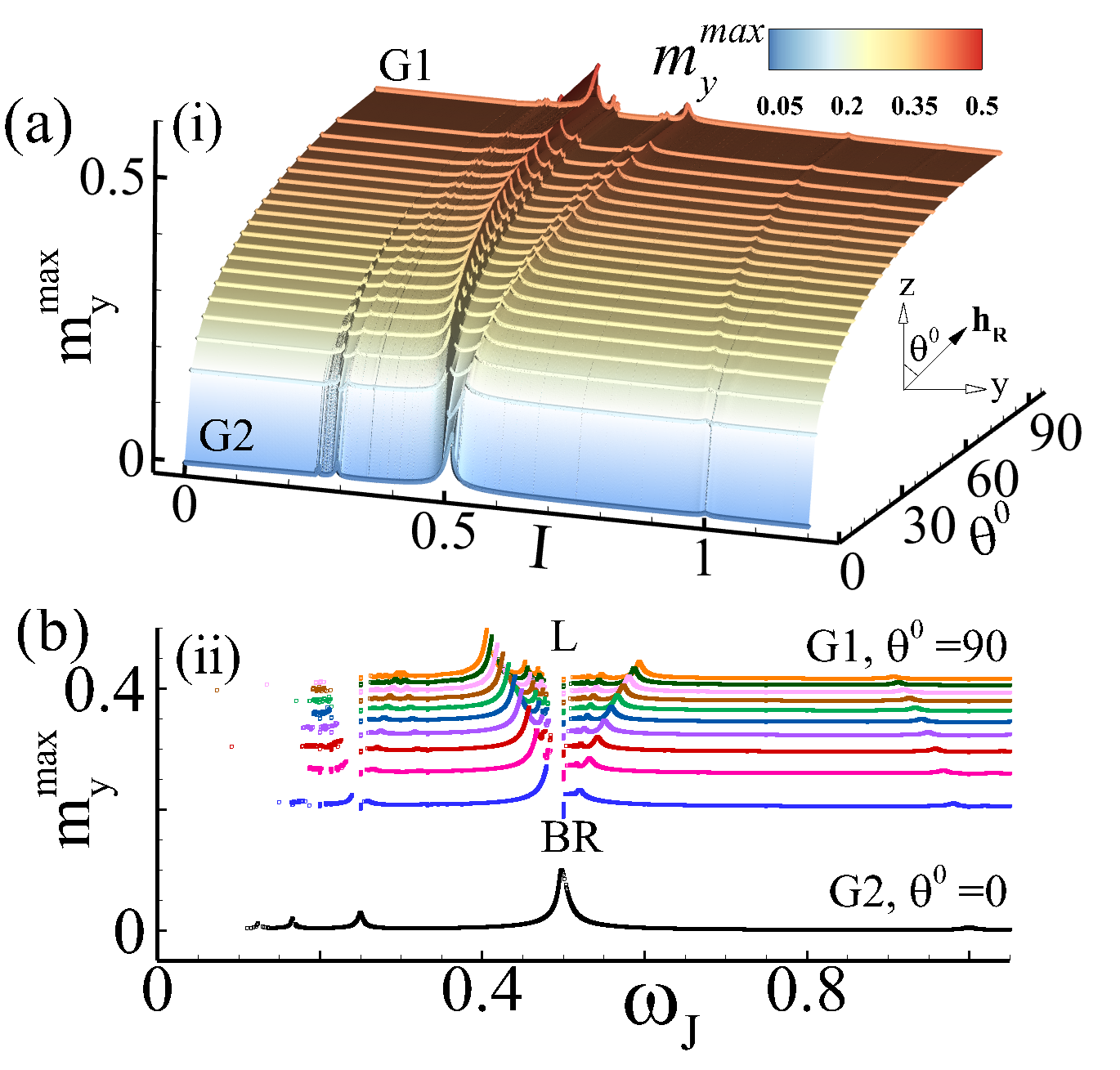}
	\includegraphics[width=0.4\linewidth]{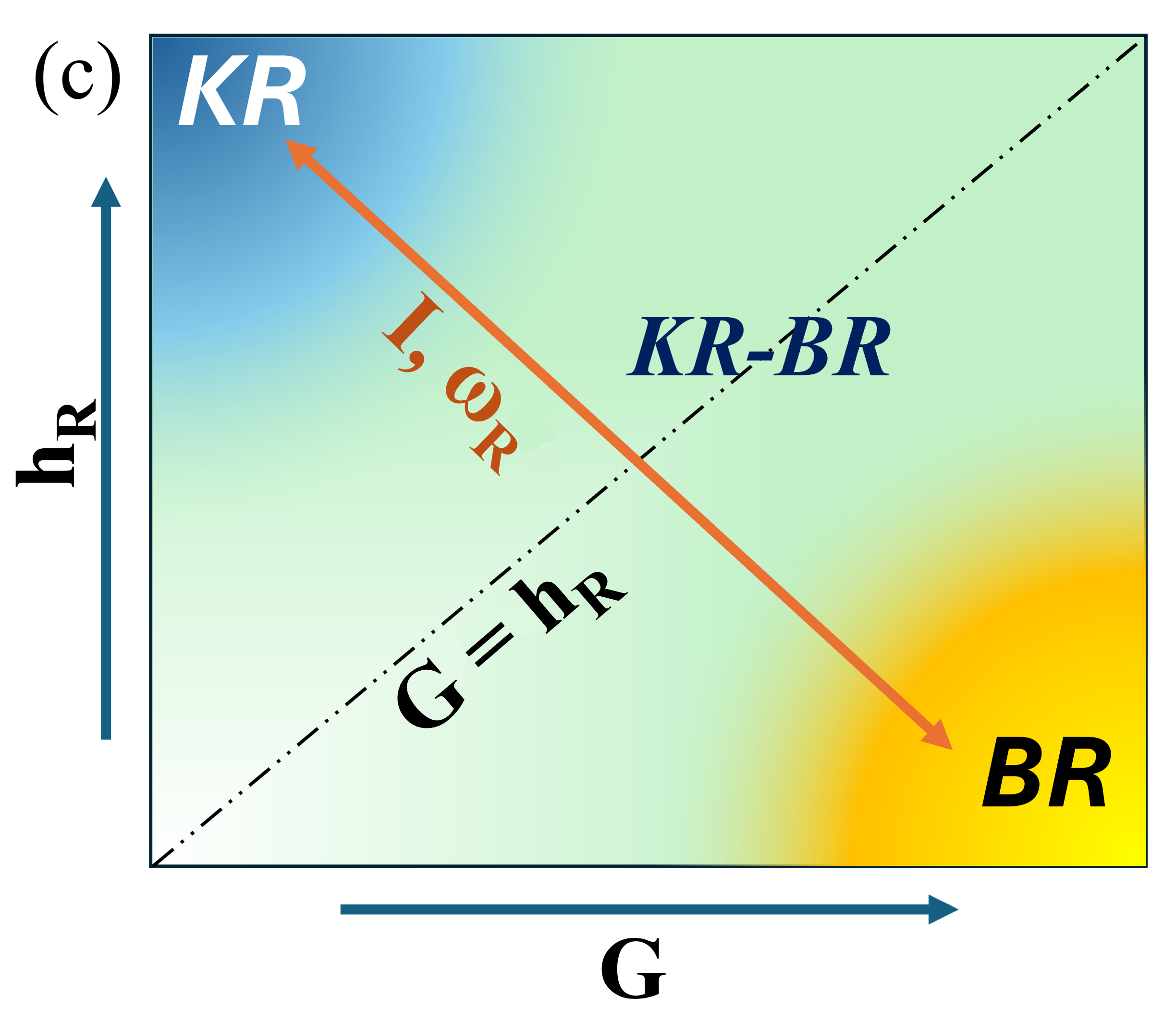}
	\caption{(a-i) Demonstration of a qualitative transformation from G1 geometry to G2 one by changing $\theta$. (b) The dependence $m_{y}^{max}$($\omega_{J}$) at different $\theta$ from $0$ to $90^{o}$ which shows the transformation from Buzdin resonance "BR" to locking "L" (Buzdin step) and appearance of combined resonances. (c) Resonance diagram for $m_{y}^{max}$ as function of radiation amplitude $h_{R}$ and the ratio of Josephson energy to magnetic energy $G$ in arbitrary unit.}
	\label{fig6}
\end{figure}	
The Buzdin resonance changes dramatically by variation of the radiation frequency $\omega_{R}$. This effect on $m_{y}^{max}(\omega_{J})$ dependence and IV-curves near $\omega_{F}$ is demonstrated in Fig.\ref{fig5}(b). In particular at $\omega_{R}=0.4$, the BR peak is 10 times smaller than at $\omega_{R}=0.2, 0.5$ and $0.6$. However, we stress that the baseline for $m_{y}^{max}$ at different $\omega_{R}$ is almost the same, which is in agreement with the fact that KR is not manifested in this geometry. Similar to G1, the Buzdin step appears at $V=\omega_{R}$. These observations are also reflected in the IV-curve. At $\omega_{R}=0.5$, the IV-curve around $\omega_{F}$ shows the BS at $V=0.5$, while the BR features are manifested at $\omega_{R}=0.6$. In supplemental material we analyze the other peaks and the corresponding IV-curve near $2\omega_{F}$ \cite{supplemental}.

\begin{widetext}
	
	\begin{table}[h]
		\centering
		\caption{Summary of resonances and locking features and the corresponding conditions in $\phi_0$ SFS Josephson junctions for G1 (in-plane) and G2 (out-of-plane) geometries. KR - Kittel resonance, BR - Buzdin resonance, BS - Buzdin step, CR - combined resonance and n, m are integers.}
		\label{t1}
		\begin{tblr}{
				colspec = {X[c,m,0.8] X[c,m,1.5] X[c,m,1.8] X[c,m,2] X[c,m,.8]},
				row{1} = {font=\bfseries},
				row{2-5} = {rowsep=0.5pt,bg=yellow!5},
				row{6-10} = {rowsep=0.5pt,bg=cyan!8},
				row{11-12} = {bg=green!5},  
				row{2-Z} = {rowsep=0.5pt},
				hlines,
				vlines,
			}
			Geometry & Phenomena & Frequency condition & Key features & Figure \\
			\SetCell[r=4]{m} G1 & KR  & $\omega_R \approx \omega_F$ & Vertical line in $I-\omega_R$ diagram; large BS in the IV-curve; dominates at $G<<h_R$ & 2(a) and S2 \\
			& BR &  $\omega_J \approx \omega_F/n$ $(n=1,2,..)$ & Set of horizontal lines in $I-\omega_R$ diagram; driven by Josephson current; resonance hump in the IV-curve; dominates at $G>>h_R$ & 2(b) and S2 \\
			& CR & $\omega_J = \omega_R \pm \omega_F/5$ & Inclined lines at KR-BR region; fractional resonance peaks ($7\omega_F/10$, $11\omega_F/10$) at $\omega_{R}=0.45$ in $m_y^{\text{max}}$ & 2(c,d) \\
			& Resonance and locking transformation & Varying $\omega_R$ near $\omega_F$ & Transformation between BS and BR in the IV-curve & 3(b) \\
			\SetCell[r=5]{m} G2 & BR & $\omega_J = \omega_F/n$ $(n=1,2,..)$ & Set of horizontal resonance lines as in G1 & 4(a,b) \\
			& KR & $\omega_R \approx \omega_F$ & Faint vertical line in $I-\omega_R$ diagram & 4(b) \\
			\SetCell[r=1]{c,m} & \SetCell[r=3]{c,m} CR & $\omega_J = (\pm m\omega_R + \omega_F)/n$ & Set of symmetric inclined lines near $\omega_{R}\approx0$ in the $I-\omega_R$ diagram & \SetCell[r=2]{c,m}4(a,b) \\
			& & $\omega_J = (\pm m\omega_R -\omega_F)/n$ & Inclined lines in the $I-\omega_R$ diagram; more pronounced by increasing $h_{R}$ & \\
			& & $\omega_J = n\omega_R/m$ & Buzdin steps in IV-curve along with resonance peaks; FMR frequency modulated by $\omega_R$ & 5(a,b) \\
			\SetCell[r=2]{m} Cross-geometry & Locking $\rightleftharpoons$ resonance & $\omega_J = n\omega_R/m$ & BS due to synchronization of Josephson and magnetization oscillations & 3(a),5(a) and 6(a,b) \\
			& Foldover $\rightleftharpoons$ resonance peak splitting &  Changing $r$ (spin-orbit coupling) and $\alpha$ (Gilbert damping) & Foldover in the BR peak in $m_{y}^{max}$ (G1); peak splitting (G2) &  3(a-iii) and 5(a-iii) \\
		\end{tblr}
	\end{table}
	
\end{widetext}

The variation of the geometry, for example, by the junction rotation, might be a way to observe the interplay of the resonance and locking effects experimentally. A qualitative demonstration of such possibility and a general diagram of the relevant resonance cases are shown in Fig.\ref{fig6}. To show the rotation effect, we consider an effective field with a transformation angle $\theta$ of the form
\begin{eqnarray}
h_{y} &=&G r \sin(\varphi - r m_{y})+  h_{R} \sin(\omega_R t) \sin\theta, \nonumber \\
h_{z} &=& m_{z}+h_{R} \sin(\omega_R t) \cos\theta,
\end{eqnarray}

where $0\leqslant\theta\leqslant 90^{o}$, and $\omega_{R}=0.5$. By changing $\theta$ the baseline for $m_{y}^{max}$ increases when the transformation from G2 geometry to G1 is realized, as can be seen in Fig.\ref{fig6}(a). This increase is a hallmark of Kittel resonance as it was shown in Fig.\ref{fig3}(b). In addition to this, the transformation from locking (L) to Buzdin resonance can be revealed clearly (see Fig.\ref{fig6}(b)). A schematic diagram of the resonance regions according to the Josephson to magnetic energy ratio and the magnetic component strength is illustrated in Fig.\ref{fig6}(c). The BR dominates when $G>>h_{R}$ (yellow region), while KR dominates in the opposite case (see blue region). At the intermediate values of $G$ and $h_{R}$, we might observe the combined resonances (green region, BR-KR), where the resonance is not determined solely by $\omega_{R}$ or $\omega_{J}$. Transformation from BR to KR and vice versa can be done by changing the bias current $I$ (changing in $\omega_{J}$) and radiation frequency $\omega_{R}$. We stress again that this transformation is the specific feature of $\varphi_{0}$ junction due to the coupling between Josephson phase and magnetization through the spin-orbit coupling. In G2 geometry the KR region suffer a strong suppression, and BR and combined resonance dominate the resonance diagram.

Finally, in Table.\ref{t1} we summarize the main features of the resonances for each geometry which are reflected on the IV-curve and $m_{y}^{max}$ diagrams with the corresponding Josephson frequency.

 \section{Conclusion}
In conclusion, we have discussed the effects which follows from the coexistence of superconductivity and magnetism in the SFS $\varphi_{0}$ Josephson junction. We have demonstrated here that the Buzdin and Kittel FMR might coexist in one system. The tuning of $\omega_{R}$ leads to the switching between these two resonances and transformation from the resonance to the locking and vice versa. Additionally, we have shown that the combined resonances can be manifested in the Buzdin and Kittel resonances intersection region. 

We would like to stress that analogous physical phenomena might appear in the system of nanomagnet coupled to the Josephson junction \cite{cai2010} which has a current-phase relationship similar to the $\varphi_{0}$ junction. In both systems an additional phase shift appears. A recent experimental work on Josephson junctions with a single magnetic atom \cite{Trahms2023} might pave the way to fabricate a $\varphi_0$ JJ weak link with NM and verify experimentally the results of the present paper. Another way would require a measuring of the magnetization state through a dc-SQUID, inductively coupled to the $\varphi_{0}$ junction \cite{Guarcello2020}. To realize the $\varphi_{0}$ state, we propose a Josephson junction with a Pt-doped permalloy ferromagnetic barrier, which provide a strong Rashba-type spin–orbit interaction \cite{Hrabec2016}. The spin–orbit interaction generates a finite $\varphi_{0}$, and measuring this phase difference as it was demonstrated in Ref.\cite{Szombati2016}, could serve as an independent way to estimate the spin orbit coupling. 

The experimental manifestation of the Kittel and Buzdin resonances can be also found by considering G1 geometry where both resonances manifest themselves depending on junction's parameters. In particular, strong difference appears when the Josephson to magnetic energy ratio is much larger that the magnitude of the radiation  magnetic component ($G>>h_{R}$) in compare with the opposite case ($G<<h_{R}$). It is demonstrated by Fig.S2 in the Supplemental material which shows two cases: (1) Kittel resonance dominates with a locking step and (2) Buzdin resonance dominates with a resonance hump instead of step.

Recent theoretical and experimental results in superconducting spintronics reveal the dissipationless charge and spin transport in SF hybrid structures by changing the texture and properties of the magnetic subsystem \cite{Buzdin2022re}. In Ref.\cite{Golubov2025} the theoretical foundations underlying a variety of phenomena in SF hybrid structures were summarized. This opens the ways for several applications as cryogenic memory \cite{Guarcello20}, superconductor–ferromagnet thermoelectric detector \cite{Geng} and superconducting spintronic tunnel diodes \cite{Narita22}. In addition, stabilizing the oscillations through the locking of Josephson and an applied ac field oscillations leads to devices that can operate across broad frequency ranges without fine-tuning and fault-tolerant superconducting logic circuits. Importantly, this can also reveal the magnetic torque-driven locking beyond Shapiro based devices which use electric field-driven junctions. Additionally, the resonant Kittel or Buzdin peaks presented without locking, provide a way to measuring the strength of spin-orbit coupling or magnetic damping by the resonance linewidth. The coexistence of resonance and locking create hybrid collective modes, where magnetic precession and charge oscillations mutually reinforce and leads to supercurrent amplification \cite{sara22,shukrinov-prb24}. 

The Shapiro step in the IV-characteristics of Josephson structures,	created by electric component of radiation, found very  wide applications in superconductor electronics and metrology.  Also, the famous Kittel resonance  is used today for different applications. We expect that Buzdin step and Busdin resonance in hybrid Josephson junctions with ferromagnetic interlayer created by magnetic component of external radiation find wide applications in the superconductor spintronics and technology. This fundamental insight into their interplay is vital for designing such  devices in the nearest future. We expect as well that the presented results make a next step in the study of the resonance and locking phenomena in anomalous Josephson junctions.\\

\section{Acknowledgments}

The authors thanks V. V. Ryazanov, I. R. Rahmonov, K. Kulikov and  D. V. Anghel for fruitful discussion. Abdelghani M. Nashaat acknowledges the financial support of the Russian Science Foundation in the framework of project 24-21-00340. Special acknowledge to Cairo university (Egypt), BLTP (Russia) within the Cooperation Agreement between ASRT, Egypt
and JINR, Russian Federation. And special thanks to the Laboratory of Information Technologies of the Joint Institute for Nuclear Research for the opportunity to use the computational resources of the HybriLIT, BLTP (JINR)  and the Bibliotheca Alexandrina’s (Egypt) for the High-Performance Computing infrastructure.

\end{document}